# Black Silicon Solar Thin-film Microcells Integrating Top Nanocone Structures for Broadband and Omnidirectional Light-Trapping

Zhida Xu[1,#], Yuan Yao[2,#], Eric P. Brueckner[2], Lanfang Li[3], Jing Jiang[1], Ralph G. Nuzzo[2,3,*] and Gang Logan Liu[1,*]

[1]Micro and Nanotechnology Laboratory, Department of Electrical and Computer Engineering, University of Illinois at Urbana-Champaign, IL 61801, USA

[2]Department of Chemistry, Fredrick Seitz Materials Research Laboratory, University of Illinois at Urbana-Champaign, IL 61801, USA

[3]Department of Materials Science and Engineering, University of Illinois at Urbana-Champaign, IL 61801, USA

#These authors contributed equally to this work

*To whom correspondence should be addressed to: r-nuzzo@illinois.edu or loganliu@illinois.edu

**Abstract**

Recently developed classes of monocrystalline silicon solar microcells (μ-cell) can be assembled into modules with characteristics (i.e., mechanically flexible forms, compact concentrator designs, and high-voltage outputs) that would be impossible to achieve using conventional, wafer-based approaches. In this paper, we describe a highly dense, uniform and non-periodic nanocone forest structure of black silicon (bSi) created on optically-thin (30 μm) μ-cells for broadband and omnidirectional light-trapping with a lithography-free and high-throughput plasma texturizing process. With optimized plasma etching conditions and a silicon nitride passivation layer, black silicon μ-cells, when embedded in a polymer waveguiding layer, display dramatic increases of as much as 65.7% in short circuit current, as compared to a bare silicon device. The conversion efficiency increases from 8% to 11.5% with a small drop in open circuit voltage and fill factor.

# 1.Introduction

With increasing power demands, solar radiation is serving a more important role as a clean and inexhaustible energy source. By the end of 2012, the milestone of 100 GW of installed photovoltaic capacity was achieved, with an annual growth rate of ~55% realized over the five year period from 2007 to 2012 [1]. In the current terrestrial photovoltaic market, silicon remains the dominant semiconductor material due to its high natural abundance, excellent optical/electronic properties and well-established production and processing routes [2]. The relatively high material cost, however, still impedes the broad use of solar cells, which counts only 0.41% of global energy consumption, ranked behind two other renewable energy sources—hydropower (3.4%) and wind power (1.39%) [3]. Solar cells based on thin films of amorphous or crystalline silicon require substantially less material compared to bulk wafer-based systems, while offering additional benefits including mechanical flexibility and ease of integration with advanced concentrating designs. One exemplary case is found in the fabrication and assembly of arrays of silicon solar thin-film microcells (μ-cell) with diverse and arbitrary spatial layouts on light-weight, flexible substrates via the transfer printing technique [4-6]. This unique design engenders advantages in allowing application-enabling distributions of devices to be made on a foreign substrate to create, as notable examples, semitransparent displays [4] or high voltage modules [7]. To compensate for the inherent low absorption of thin-film silicon, different concentrating optical components are integrated with μ-cells, including micro-lens arrays [4], nanostructured backside reflector [8] as well as luminescent waveguides [9]. The more common strategies used to provide enhanced light absorption in commercial cells, surface texturation (e.g. alkaline etching [10] and micromachining [11]), cannot be easily accommodated in the fabrication of μ-cells. Efforts have been made to reduce the top surface reflection on μ-cells by adding a single-layer anti-reflection coating (ARC) [12] or creating nano-pillar arrays patterned by soft imprint lithography [13]. Both methods have their limitations. The ARC layer can only suppress the reflection for a limited range of wavelengths and angles. The nano-pillar array achieves broader wavelength coverage, however, its pillar-like geometry is not optimized to minimize reflections according to effective medium theory (EMT) [14], as the effective refractive index (RI) change within the structure is rather abrupt.

The black silicon (bSi) process, which is capable of creating nanocone structures with a gradual RI change at the silicon/air interface, is an alternative solution which can potentially further improve the broadband and omnidirectional absorption of a μ-cell device format and enable the full use of solar energy over the whole spectrum. Black silicon is a semiconductor material whose surface is modified with micro- or nanostructures in ways that make it highly absorptive and thus appear black. It was discovered in the 1980s as an unexpected side effect of the use of reactive ion etching (RIE) in the semiconductor industry [15]. Over the years, its potential has been recognized in various research areas including superhydrophobicity, ARCs for photovoltaics, highly sensitive photodetectors in optoelectronics, and biomedical sensing, for this reason, motivating intensive efforts to provide new and/or optimized approaches to the production of bSi materials [16-18]. In addition to RIE, effective processing methods to produce bSi include metal-assisted chemical etching [19-21] and pulsed-laser irradiation [22]. Among them, though, RIE retains the advantages of high throughput, low cost and short manufacturing cycle. We have developed a 3-step lithography-free RIE process (15~20 minutes per cycle) that is capable of producing wafer-scale bSi with dense nanocone forests on silicon surfaces of various doping type, crystallographic planes and morphologies [23]—structures that in past work have been applied to both optical sensors [24] and solar cells [25]. The adaptability of this RIE process enables the creation of bSi on photovoltaic devices after doping, circumventing the deleterious consequences of uneven doping that tend to degrade the performance of nanostructured silicon materials [20]. It was demonstrated by us previously that the efficiency of a bulk silicon solar cell could be increased by 14.7% using this RIE-based bSi process [25]. The mask-free nature of this approach also enables a simple means through which textures can be added to μ-cells to mitigate losses from top surface reflections and further increase their power output. In this work, subwavelength bSi nanocone structures are produced on μ-cell devices (black μ-cell) for broadband and omnidirectional light trapping. Combined with a silicon nitride passivation layer [26], the energy conversion efficiency ($\eta$) of the black μ-cell embedded in a polymer matrix increases by 42.8%, and short circuit current ($I_{sc}$) increases by 65.7%.

## 2. Production of bSi on μ-cell

The bSi was produced with a 3-step lithography-free plasma RIE process in a PlasmaTherm SLR-770 inductively coupled plasma reactive ion etcher (ICP-RIE), reported in details by us previously [23]. Fig. 1 shows the cross-sectional schematic of the bSi fabrication process. First, a thin layer of oxide was formed on the silicon surface under an oxygen plasma for 5 min (Fig. 1(a)). The oxide was etched using a high-power $CHF_3$ plasma to form randomly dispersed oxide islands (Fig. 1(b)). Finally, a HBr plasma was used for the silicon etching step. The random oxide mask protects the silicon underneath it from being etched by HBr plasma and the nanocone forest is in this way created (Fig. 1(c)). The oxide mask was removed using a buffered oxide etchant (BOE), leaving uniform and large-area bSi with dense nanocone structures on the silicon surface, as shown by both the optical and SEM images in Fig. 1(d) and 1(e). The tapered shape of sub-wavelength structures created by this approach makes the effective refractive index gradually increase from the top (air) to the bottom (silicon) of the nanocone forest without abrupt change. As predicted by the effective medium theory (inset on the right of Fig. 1(e)) [14,27], the reflection is greatly suppressed and the omnidirectional absorption enhanced over both the optical [28,29] and radio frequency ranges [30]. Compared to metal-assisted wet etching methods [19-21], our RIE process offers several benefits: (1) both mask creation (step 2) and etching (step 3) are completed in the same plasma chamber in a continuous fashion in about 15~20 min; (2) the process eliminates the need of using metal, a material usually disfavored in a semiconductor processing route due to potential contamination issues; and (3) the anisotropic nature of the RIE dry etching makes it better suited than the isotropic wet chemical etching as a means for creating high aspect-ratio silicon nanocones.

The advantages of this 3-step RIE process lie in its high controllability compared with other RIE bSi methods. By altering the conditions in each step, different properties of the nanocone forest can be tuned effectively: First, the tapered nanocone shape is mainly determined by the ratio between the RIE DC bias (which controls directional physical ion bombardment) and the inductively coupled plasma (ICP) power (which controls the plasma density and isotropic radical reaction) in step 3. A higher DC bias will lead to nanocones with straighter sidewalls, while a higher ICP power would result in a more tapered morphology. Second, the density of the nanocones can be controlled by the power and etching time used in step 2, which uses a $CHF_3$ plasma to randomly etch the oxide layer to form the required oxide nanomask. We can reliably attain the highest density structures with a 300 W power setting for the RIE and 500 W for the ICP with an etching time of 2 min (step 2). Third, the depth of the nanocones can be changed by varying the HBr etching time (step 3). In the supplementary materials, we included the SEM images (Fig. S1) and absorption spectra (Fig. S2) of bSi attained with different set values of etching time (from 1 min to 10 min). It is shown in these SEM images (Fig. S1) that the length and width of the nanocones initially increase with a longer etching duration (up to 6 min), corresponding to an associated higher level of optical absorption (Fig. S2). For the case of longer etching-time values of 10 min or more, the nanocone structures attained are finer in scale, with defects and even broken nanostructures evidenced that are further correlated with degraded optical properties. The bSi samples attained using an intermediate value of the etching time, here of 6 min and 8 min, display both similar morphology and optical absorption. We chose 6 min as the optimal processing condition for bSi devices for the reason that longer etching induces more defects on the surface.

The μ-cells (30 μm thick, 100 μm wide and 1.5 mm long) utilized in this work were fabricated on a p-type Czochralski silicon wafer (i.e. donor wafer) by a previously reported process of photolithography, deep ICP-RIE etching and undercut in basic anisotropic etchant solutions (the oxide mask was removed by BOE for the bSi etching) [12]. Fig. 2(a) illustrates the structure of the device with a vertical contact scheme, highlighting the phosphorus- (n+, green), and boron- (p+, blue), and intrinsically (gray) doped regions. Following the completion of the device fabrication, the μ-cells with top contact pads (gold) are amenable for transfer-assembly onto secondary substrates bearing pre-patterned bottom contact bus lines in programmable layouts, as shown by the SEM image presented in Fig. 2(b). The devices were then planarized with a photocurable liquid resin (NOA61, Norland Products, Inc), leaving μ-cells embedded in a polymer matrix before bSi etching (Fig. 2(c), orange region is the reflective silicon surface, the gold square and line are the top and bottom contacts for the device). The RIE process described earlier was utilized to create dense and sharp nanocone features on top of the devices, rendering them black (Fig. 2(d), the polymer matrix was also textured by the RIE, making the background darker than that seen in Fig. 2(c)). The depth of the nanocone surface texture was controlled to fall in the range of ~200 nm (Fig. 1(e)) by tuning the etching time (6 min). Although the absorption enhancements from bSi generally increase with the feature size, the nanocone depth needs to be tailored to avoid excessive surface defects and the penetration of the p-n junction (less than 1 μm deep [12]) of the photovoltaic device. (It is shown by Fig. S3 in the Supplementary Materials that after 6 min etching, the current output of the device begins to reach an asymptotic limit; the dark saturation current of the device increases as well with the use of longer etching time.) To passivate the dangling bonds of the black silicon surface and thus reduce the surface recombination rate, we deposited 20 nm of $SiN_x$ onto the black μ-cells by plasma enhanced chemical vapor deposition (PECVD, PlasmaLab) at 220 °C, which comfortably covers the nanocone features (see the SEM images before and after deposition in Fig. 2(e) and 2(f)).

**3. Angular Absorption Spectra of bSi**

The broadband and omnidirectional light trapping properties of bSi on devices were examined by measuring the angular absorption spectra of a bulk bSi wafer, processed using RIE conditions identical to those applied to the μ-cells. This measurement was made with a Cary 5000 UV-Vis-NIR spectrophotometer equipped with an integration sphere, as illustrated in Fig. 3(a). The sample, mounted in the center of the sphere, was rotated along its z axis to vary the incidence angle ($\theta$, relative to the norm of the wafer) of the incoming monochromatic light beam along the x axis. The reflected and scattered light not absorbed by the silicon surface is collected by the photo-detector at the bottom, the signal from which was subtracted from that measured for 100% reflection (calibrated using a diffuse reflector) to determine the angle-dependent light absorption of the sample. The values of effective absorption ($A_{\text{eff}}$) measured between 300 and 1100 nm at each angle, as calculated from these data were tabulated in Table S1 in the supplementary materials and plotted in Fig. 4. The $A_{\text{eff}}$ values were calculated by using the $R_{\text{eff}}$ method, taking into account the solar flux distribution found under AM-1.5G standard solar irradiation [31]. The absorption spectra of polished p-type silicon samples (identical to the wafer used for cell fabrication) measured before and after the RIE bSi processing are presented in Fig. 3(b) and 3(c), respectively. The bSi nanostructures dramatically enhance the absorption by more than 30% over the whole measured spectral range, here from 300 nm to 1100 nm (in the UV region, the enhancement is even greater—between 50% and 70%). The bSi sample shows a significantly higher absorption of the light than a polished one even at high incident angles ($\theta > 60°$), where the absorption of the latter declines rapidly with increasing $\theta$. At $\theta = 70°$, for example, bSi still achieves an absorption of over 90% between 300 nm and 800 nm, while that of the original silicon drops below 50%. Depositing 20 nm of $SiN_x$ passivation layer onto the bSi wafer (compare Fig. 3(d) with Fig. 3(c)), reduces its ability to absorb incident light for $\theta > 60°$, but as seen in the data still outperforms the polished sample over a broad wavelength range.

## 4. *I-V* Characteristics of black μ-cell

The effects of the subwavelength nanocone features on the photovoltaic performance of the μ-cells were investigated by measuring their *I-V* characteristics under a simulated AM1.5G solar spectrum of 1000 W/m$^2$ at room temperature. Fig. 5 shows *I-V* curves of the same μ-cell after different processing steps, including: transfer-printing (original cell, bare silicon); embedding in a polymer matrix (after planarization); RIE treatment; and SiN$_x$ deposition. A non-reflective anodized aluminum plate was placed under the device array in all cases to suppress backside reflections. The waveguiding effects [12] from the polymer matrix lead to a 23.3% increase (from 31.37 to 38.68 μA) in current output from a representative cell (as seen comparing the black and red curves in Fig. 5). RIE treatment adds another 6.76 μA enhancement to the short circuit current (green curve in Fig.5) due to the light trapping properties of the top nanocone structures, bringing the total relative enhancement to 44.9%. As compared with the original device, however, both the open circuit voltage ($V_{oc}$, decreases from 0.515 V to 0.485 V) and fill factor (FF, decreases from 0.712 to 0.622) deteriorate, as the RIE also generates surface defects during the etching process. A 20 nm thick PECVD SiN$_x$, deposited as a passivation layer, partially repairs the dangling bonds on the surface, resulting a significant enhancement in $I_{sc}$ (from 45.44 to 51.95 μA, as seen by comparing the green and blue curves in Fig.5) and a slight recovery of the FF (from 0.622 to 0.645) and $V_{oc}$ (from 0.485 to 0.490 V). After all the processing steps, the short circuit current of the device increases dramatically by 65.7% (compare black and blue curves in Fig. 5), a result obtained as a consequence of enhanced photon capturing properties introduced by the bSi nanostructures. Relative to the original device, the efficiency ($\eta$, calculated based on the top surface area of the device without accounting for contributions from light incident on the side of the cells) improves by 42.8% (from 8.07% to 11.52%) after $SiN_x$ deposition, a value principally limited by the increased surface recombination rate induced by the RIE process.

## 5. Discussion

The continuously tapered morphology of the bSi nanostructures demonstrated in this work bears a close resemblance to the pyramidal subwavelength pillar arrays found on moth eyes [32]. These natural structures provide a superior graded index profile at the air/device interface that greatly suppresses top surface reflections. As compared to cylindrical pillars, a more common fabricated structure that mainly diffracts light into µ-cells [13,33], the non-periodical nanocone forests created by the lithography-free RIE step elicits a different photon capturing mechanism, one characterized by photon randomization or scattering that is largely insensitive to its wavelength or incident angle. These broadband and omnidirectional absorption enhancements translate directly into boosts in power output of as much as 42.8%, leading to a more efficient utilization of the active photovoltaic material (silicon). The power-referenced material consumption here could be significantly lowered to further improve the system's cost-effectiveness when coupling devices with compact concentrator designs, including microlens array [4] or luminescent waveguides [9] as demonstrated in our earlier studies of concentrator designs using flat µ-cells. The concentrated photon flux would be more effectively converted to electrical power by the black µ-cell than a device without texturing due to the increased optical interaction length within the nanocone structures.

As evidenced from an inspection of the experimental results presented above, however, the relative enhancement for the energy conversion efficiency (~40%) of the black silicon μ-cell is not as significant as that for the absorption or short circuit current values (~60%), as the carrier recombination losses introduced by the RIE etching is only partially mitigated by the $SiN_x$ passivation layer. Modifications might be made to the RIE process to reduce the processing-generated surface defects. One possible source of defects is the severe ion bombardment that accompanies the HBr etching in the third step, a result of the high atomic weight of bromine ion (atomic weight = 80 Da). The ion related energy transfer can be reduced by using $Cl_2$ as the etching gas. Since the chlorine ion has a much lower atomic weight (35.5 Da), the etching process would be dominated by chemical reactions with chlorine radicals rather than physical bombardment, leading to less surface defects albeit also engendering a slower etching rate. The etching protocols, however, can be adjusted in principle to generate the same nanocone structures as are obtained with the HBr process.

PECVD deposited dielectrics (including $SiN_x$ and hydrogenated amorphous silicon (a-Si:H)) have been shown to effectively passivate the surface of macro-scale solar cells due to the embedded hydrogen atoms at the Si/dielectric interface [26,34]. For a black μ-cell with a highly textured surface, however, processes with improved step coverage and film quality, such as atomic layer deposition (ALD) [35] and thermal oxidation [12,20], could potentially generate a passivation layer with better uniformity and lower surface recombination velocity. Hydrogen passivation is an alternative approach that in principle is capable of repairing deeper dangling bonds due to the longer diffusion length of the hydrogen atom [36]. The high temperature requirements (>500 °C) of these above-mentioned processes, however, are not compatible with transfer-printed μ-cells, as both supporting elements (i.e. polymer matrix, glass substrate) and metal contacts would be damaged. Significant modifications would need to be made to the μ-cell fabrication sequence to incorporate the b-Si etching and high-temperature passivation steps before metal contact patterning, which is both a challenging and interesting direction to pursue in the future.

## 6. Conclusion

We produced black silicon (bSi) nanocone structures on fully functional silicon solar thin-film microcells (μ-cells) with a simple 3-step RIE process along with silicon nitride passivation layer for broadband and omnidirectional light absorption enhancements. The photon capturing properties of the black μ-cell increased significantly, as shown by the marked short circuit current enhancements of as much as 65.7%. The RIE process, however, also creates surface defects in the device, which limits the enhancements in power output (~40%). The increased surface recombination could be potentially mitigated by both advanced passivation methods (e.g. ALD $Al_2O_3$, thermal oxidation, hydrogen passivation) and adjustments to the etching process. In addition, optical concentration elements (e.g. backside reflectors, microlens array and luminescent waveguides) and module assembly strategies developed for microscale devices in our previous work [4,9,12] could also be utilized for the black µ-cells to create large-scale functional arrays of high power output with low material consumption [9].

**Acknowledgement**

This work is supported in part by a NSF grant ECCS 10-28568 and DOE grants DE-SC0001293 (subcontract no. 67N-1087758) and DE-FG02-07ER46471.

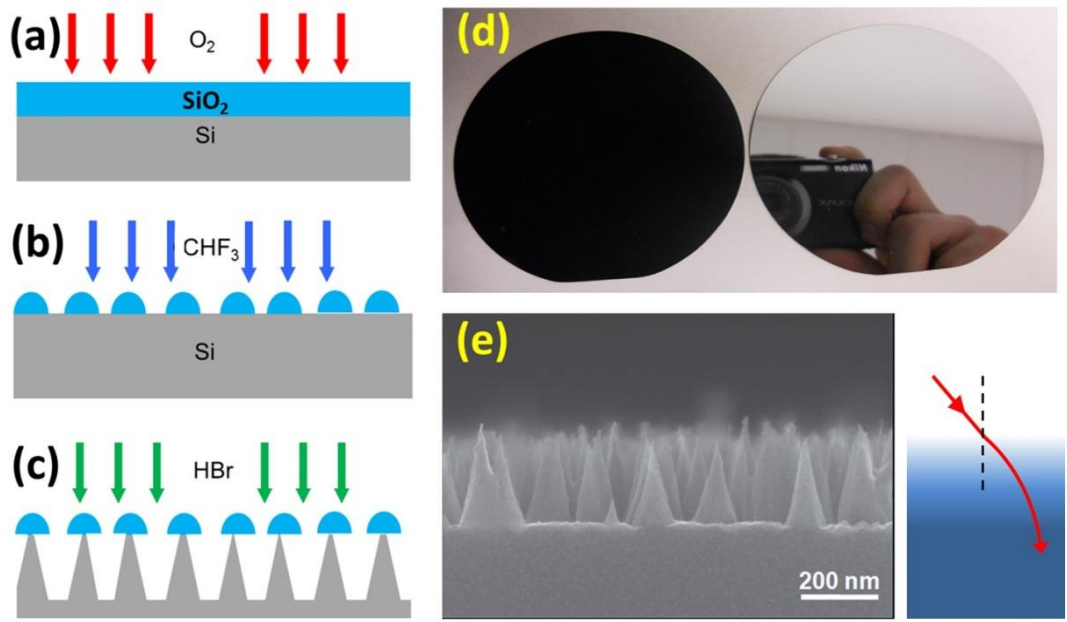

Figure 1 (a-c) Fabrication process of bSi. (a) Thin oxide layer on silicon surface formed by oxygen plasma. (b) Randomly dispersed oxide mask formed by $CHF_3$ plasma etching of thin oxide layer. (c)Silicon nanocones formed by HBr plasma etching. (d) Comparison of a 3" polished silicon wafer (right) and a 3" bSi wafer. (e) Cross-sectional SEM image of nanocone forest on bSi. The inset on the right shows how the gradient effective refractive index of nanocone forest enhances the absorption.

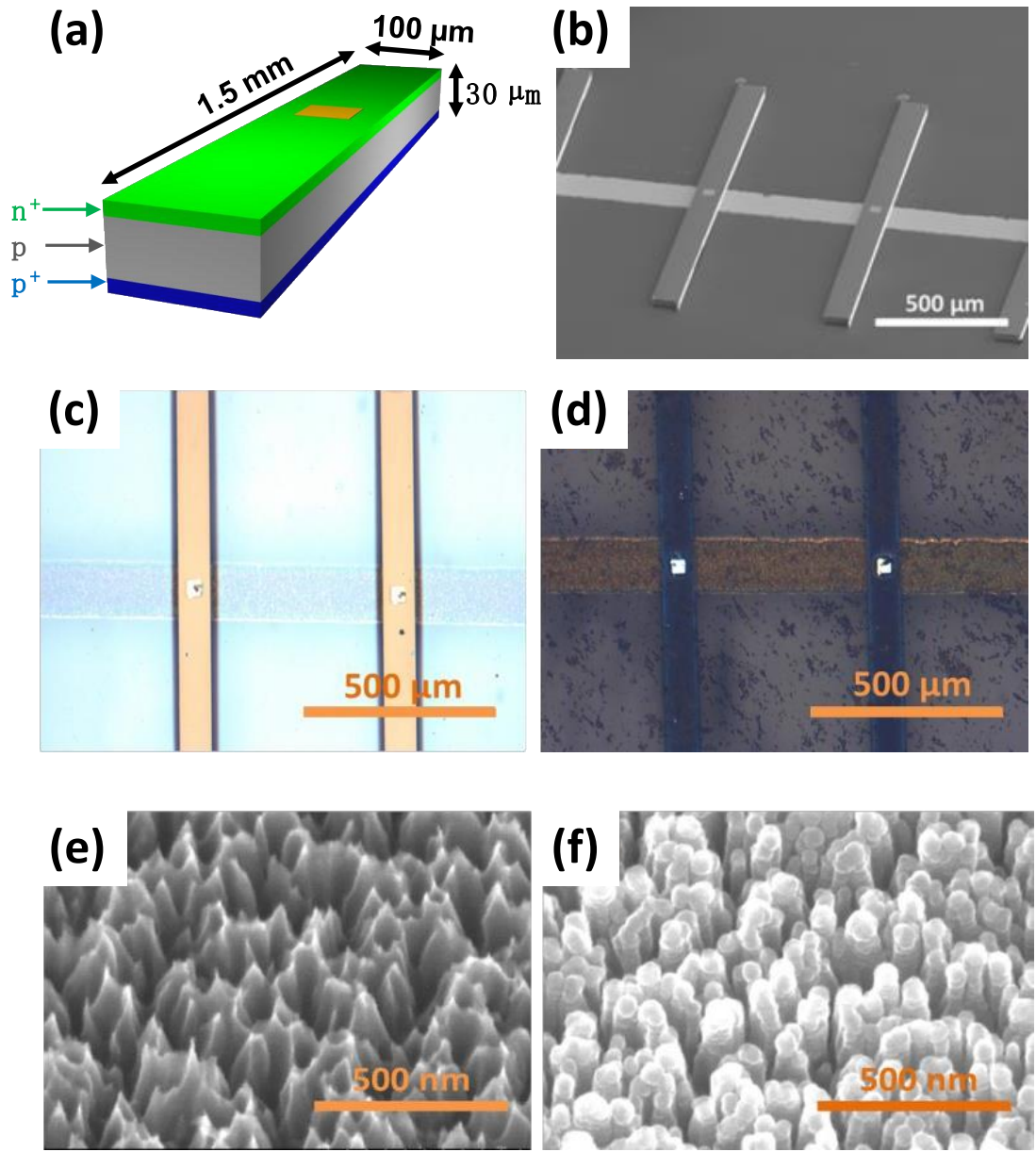

Figure 2 (a) Schematic illustration of the structure of a silicon solar microcell. (b) A SEM image of an array of μ-cells after being transfer-printed onto glass substrate finished with top contact pad and backside contact bus line. (c) Optical image of original μ-cells embedded in a polymer matrix. (d) Optical image of black μ-cells treated by the 3-step RIE process. (e) SEM image of the surface of a black μ-cell. (f) SEM image of the surface of a black μ-cell after deposition of 20 nm $SiN_x$.

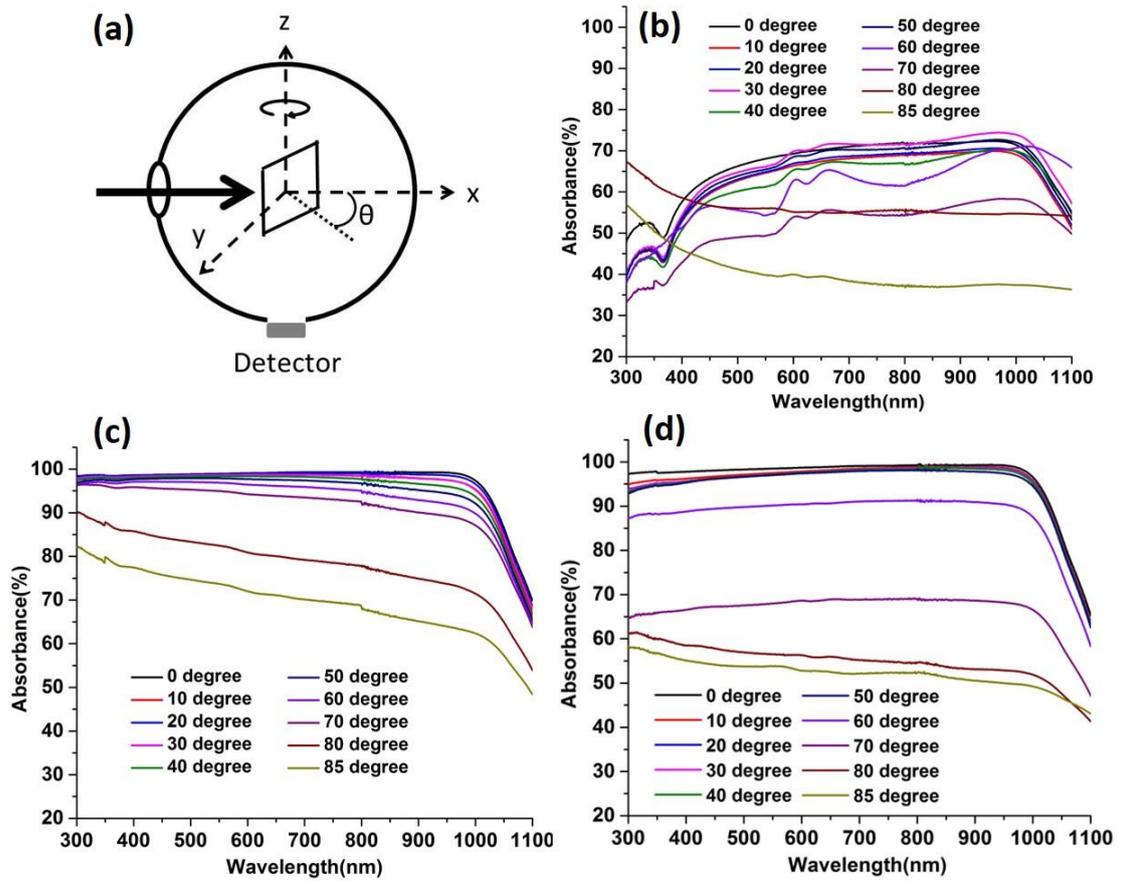

Figure 3 (a) Illustration of the experimental setup of integration sphere measurement for absorption spectra of bSi. $\theta$ is the angle between the incident light beam and the norm of sample. (b) Absorption spectra of the polished p-type silicon wafer at $\theta$ from 0º to 85º. (c) Absorption spectra of p-type bSi wafer at $\theta$ from 0º to 85º.(d) Absorption spectra of p-type bSi wafer deposited with 20 nm of $SiN_x$ at $\theta$ from 0º to 85º.

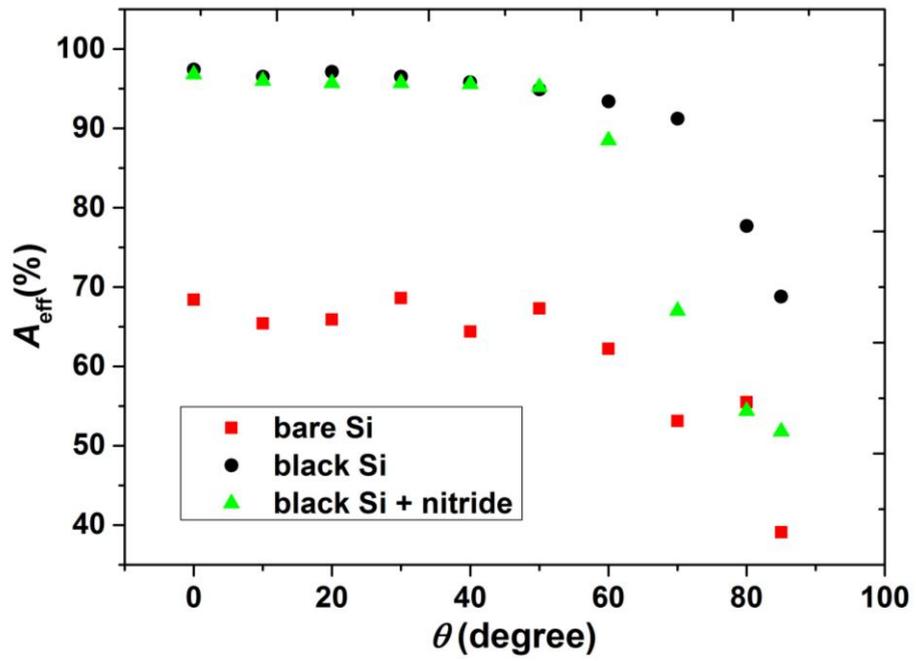

Figure 4 Effective Absorption ($A_{eff}$/%) of bare Si, bSi and bSi+nitride at different incident angle $\theta$.

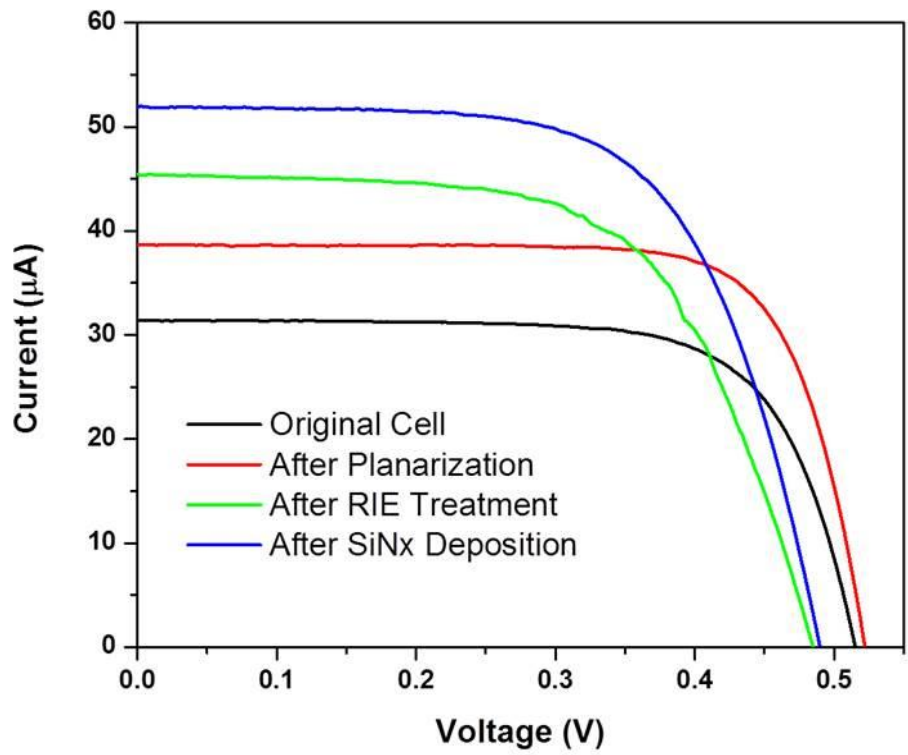

Figure 5 *I-V* characteristics of the μ-cell under AM 1.5G solar spectrum after different fabrication steps